\documentclass[12pt,oneside]{amsart}

\usepackage{url}

\newcommand{\eg}{{\it e.g.\/}\ }

\newcommand{\nil}[1]{}

\begin{document}

\thispagestyle{empty}

\begin{center}
{\large\bf A SIMPLE E-VOTING PROTOCOL}
\end{center}

\vspace{0.8cm}

\begin{center}
Fr\'ed\'eric Connes\footnote{Doctoral student,
  Universit\'e Paris-II Panth\'eon-Assas, Paris, France.}\\
{\small\url{frederic@connes.org}}\\
\bigskip
August 18, 2008
\end{center}

\vspace{0.8cm}

\section*{Abstract}

\scriptsize\bfseries
We propose an e-voting protocol that seems to allow citizens
to verify that their vote has been accurately taken into account
while preserving its secrecy, without requiring the use of a complex
process. The main idea is to give each voter a receipt on which her choice is
mixed with the choices of other voters.
\normalsize\normalfont

\vspace{0.8cm}

\section{Introduction}

As shown by many studies since the beginning of the years 2000
(\eg~\cite{Koh04}), e-voting systems do not offer a sufficient
security level. In particular, they do not guarantee the exactness of results
while at the same time protecting the secrecy of votes.

The most commonly used solution nowadays is to print a paper
ballot~\cite{Mer92} so that voters can check that the machine registered their
choice accurately. The paper ballots can then be manually counted, preferably
on a random basis, which gives a result that is independent of the electronic
result and has the same reliability as the old manual way of counting votes.
A problem can thus be detected and corrected. However, the major drawback of
this approach is that it is not possible to manually recount the ballots all
the time, since it would render the electronic counting useless.
Moreover, the manual counting is not necessarily easy to perform, notably in
countries like the U.S. where many elections take place on the same day.
It should also be noted that the blinds cannot
check by themselves that their vote has been correctly printed.

To solve these problems, several protocols aiming at giving a receipt to the
voters have been developed (see \eg~\cite{Cha04}, \cite{Riv0610},
\cite{Cos08} and \cite{Cha08}). They have the great advantage of allowing
citizens to check that their vote has been properly taken into account, which
is reassuring. Each vote can be verified, wherever it has
been cast. Furthermore, receipts
do not require any manual recounting, each citizen being responsible for
the exactness of their vote. However, these protocols have the drawback of
being complex, because the receipt must not be a proof of the vote, otherwise
the secrecy would not be protected. Some protocols which do not make use
of cryptography have been proposed by Ronald~L. Rivest and
Warren~D. Smith~\cite{Riv07}, but they are based on a rather complex way of
voting or on the
generation of a receipt corresponding to another citizen, which does not allow
voters to check that their own vote has been taken into account.

Here, we propose a voting protocol that seems to guarantee the integrity
of votes while preserving the secrecy,
without requiring the use of a complex process. This
paper aims to initiate a reflection on this protocol, which certainly
remains perfectible.

\section{Principle}

We think that the only solution to fully ensure the exactness
of the results is to make the choices of the voters available publicly,
without of course publishing the names of the voters.
For instance, the choices could be anonymously published on a Web site,
in such a way that the citizens could tally the votes and
verify that their own vote has been accurately taken into account.

To achieve this goal, we propose to assign a unique
ID, which is just a random number, to each voter
and to only make the correspondence between the
IDs and the choices public. If the correspondence between the
IDs and the voters remains secret, then the publication of the votes
becomes anonymous. It is thus essential to find a way to preserve at all price
the anonymity of the IDs. This means that the link
between the voter and her ID must not survive outside the
polling booth. Then, only the voter will know the ID associated with
her vote.

To put this into practice,
one can imagine that, after the validation of the choices, the machine
displays an ID that the voter will memorize. However, this
approach is flawed. That is because the voter only knows one valid ID, which is
the one attributed to him by the machine. This prevents the citizen from faking
having voted for someone else, since he has no way to invent a correct ID
corresponding to another choice.
Moreover, with such a system, the voter cannot have the proof that the
ID that is attributed to him is not identical to the ID that was
attributed to another voter having made the same choice before.
Then, there would be one spare ID, which the machine could use to create
a fraudulent vote. There would be no way to detect such an attack.
Finally, the envisioned system has the major drawback of not providing
the voter with a receipt allowing him to complain to a court in case his choice
was not correctly taken into account.

To solve these problems, we first propose that the ID be
displayed to the voter just after he enters the polling booth. At
this point, the machine cannot know what the choice of the voter will be.
If the ID is displayed from the beginning, it will be a new ID since
otherwise the machine would have to bet on the choice that will be made.
If it picks up an ID that eventually does not match the choice of the voter,
the fraud would be discovered. However, this does not completely remove the
risk of cheating, as the machine could bet and win.
In this respect, it should be noted that it is not possible to
let the voter choose his ID, because it could be used as a sign of recognition.

Now, in order to preserve the secrecy of votes and to provide the voter with a
useful receipt, we propose that the machine prints a piece of paper that links
each available choice (candidates or answers) to an anonymous ID, all the
pairings being valid : the actual choice of the voter would be linked to her
assigned ID, and each other choice would be linked to the ID of a citizen who
previously voted for this choice. These previously used IDs would be taken
randomly. Consequently, when the voter leaves the polling booth with her
receipt, the latter contains all available choices linked to anonymous IDs, all
the pairings being valid and looking equal on the face of it. Thus, it will be
impossible for a third person to find, among these pairings, which one
corresponds to the actual choice of the voter. But the voter will know her ID,
since it is written in front of her choice. At the end of the day, the voter
can go to the Web site that publishes the votes and check that her choice has
properly been taken into account. She can also verify the other pairings on her
receipt, and if she detects a problem, she can go to a court with her receipt,
which is a proof allowing her to ask for a correction of the results. The court
will have no way to find out if the complaint concerns the vote of the
plaintiff or that of another person.

To avoid forgery, it seems mandatory that the receipt contains a
digital signature, which will allow the court to control its origin.
It could also be printed on a paper difficult to counterfeit.

\section{Example}

We now give an example of the proposed protocol.

Let an election bring together 4 candidates: A, B, C and D.

When the voter enters the polling booth, he activates the
machine, which displays the ID ``1597362523648''. If the voter is
blind, the ID is listened on headphones.

The voter then makes a choice, for instance C.

The machine prints a receipt, as shown in figure~1.

\begin{center}
\begin{figure}
\ttfamily
\begin{tabular}{|c c|}
  \hline
  & \\
  \multicolumn{2}{|c|}{Presidential Election} \\
  \multicolumn{2}{|c|}{November 4, 2008} \\
  \multicolumn{2}{|c|}{Foo County, Bar State} \\
  & \\
  \ \ A & 6597853518467 \ \\
  \ \ B & 9431587321355 \ \\
  \ \ C & 1597362523648 \ \\
  \ \ D & 3943873165496 \ \\
  & \\
  \multicolumn{2}{|c|}{Signature:} \\
  \multicolumn{2}{|c|}{A31I321H54B87Q3} \\
  \multicolumn{2}{|c|}{F4W654P98G321C6} \\
  \multicolumn{2}{|c|}{84N9835Z1684H35} \\
  \multicolumn{2}{|c|}{B6F9687P35I56AP} \\
  & \\
  \hline
\end{tabular}
\rmfamily \caption{Voter Receipt}
\end{figure}
\end{center}

One can see that a previous voter with the ID ``6597853518467''
voted for A. The voter with the ID ``9431587321355'' voted for B,
and the voter with the ID ``3943873165496'' voted for D. Our
voter ignores the identities corresponding to these IDs.

The receipt is displayed behind a glass, so that the voter can check it without
touching it.
If the voter is blind, he can ask someone to read the receipt to him.
The help will ignore where the actual vote is, but the blind will be
able to recognize his ID. In that way, the secrecy will be protected despite
the presence of the help.

If the receipt is incorrect, the voter cancels his vote and the machine
immediately destroys the receipt without letting the voter touch it.
Otherwise, the voter validates his choice and he can take the receipt.
The machine is then initialized, and the voter can leave the polling booth
with his receipt. This procedure ensures that the voter only gets one receipt.

At the end of the day, the voter can go to
the Web site publishing the votes and find the data shown in figure~2.

\begin{center}
\begin{figure}
\begin{tabular}{|c r|}
  \hline
  & \\
  \multicolumn{2}{|c|}{Presidential Election} \\
  \multicolumn{2}{|c|}{November 4, 2008} \\
  \multicolumn{2}{|c|}{Foo County, Bar State} \\
  & \\
  \hline
  & \\
  \multicolumn{2}{|c|}{\textit{Votes:}} \\
  & \\
  \multicolumn{2}{|c|}{$\cdots$} \\
  A & 5231897463515 \\
  A & 6597853518467 \\
  A & 8795462163516 \\
  \multicolumn{2}{|c|}{$\cdots$} \\
  B & 4546138496616 \\
  B & 7894611685366 \\
  B & 9431587321355 \\
  \multicolumn{2}{|c|}{$\cdots$} \\
  C & 1597362523648 \\
  C & 2923578356914 \\
  C & 7898756465486 \\
  \multicolumn{2}{|c|}{$\cdots$} \\
  D & 3943873165496 \\
  D & 4567315796865 \\
  D & 7986543546933 \\
  \multicolumn{2}{|c|}{$\cdots$} \\
  & \\
  \hline
  & \\
  \multicolumn{2}{|c|}{\textit{Results:}} \\
  & \\
  A & 1863 \\
  B & 536 \\
  C & 2013 \\
  D & 289 \\
  & \\
  \hline
\end{tabular}
\caption{Results Web site}
\end{figure}
\end{center}

The voter can check that the votes have correctly been taken into account.
He can also tally the votes.

To prevent the addition of fraudulent votes, the number of signings
needs to be publicly counted and displayed on the Web site.
Ideally, the name of the persons who voted would also be published on a Web
site, so that the number of voters could be verified by anyone.

It should be noted that if the law permits the voter to choose more than one
candidate, the machine should provide as many IDs as the number of possible
choices, each ID then being linked to a single choice.

\section{Bootstrap}

The main problem posed by this protocol is that
when the first voter casts her ballot, there is no valid ID corresponding to
the other choices, for the simple reason that there are no previous voters.
We thus face a bootstrap problem.

In order to overcome this difficulty, we propose that, at the start of the day,
the machine generates a certain number of votes for the
various candidates. For instance, each candidate would receive 10~votes,
linked to IDs corresponding to fake voters. These 10~votes
would be published on the Web site, but nobody would be able to distinguish
them from the votes corresponding to actual voters.
In fact, only the first voter could spot these votes, but he could not
prove to a third person that they are fake.
Of course, at the end of the day, one needs to remove 10~votes from each
candidate's figures to obtain the correct result.

This approach poses a problem if the machine cheats and does not generate
10~bootstrap votes for each candidate.
For instance, it could generate 5~votes for A and 15 for B. Nobody
would be able to detect this fraud if A receives at least 6 actual votes, since
the total would be greater or equal to 11, which means that the final result
would be greater or equal to 1. This way, the citizens having voted for A would
think that this vote is theirs, and they would not complain.

To solve this problem, we propose that as soon as the machine has generated the
bootstrap votes it records them on a removable media while signing them with a
private key and encrypting them with the public key of a trusted authority,
such as the electoral commission or a court. These encrypted votes could then
be transmitted, for instance via the Internet, to the trusted authority, which
would decipher them and check the signature. It would control that the
bootstrap votes have been correctly generated and, at the end of the day, would
verify that these votes actually appear on the Web site. All these operations
can be performed manually, but they can also be automated. The only point one
needs to be careful about here is that the trusted authority be distinct from
the court in front of which the complaints of voters are dealt with. Otherwise,
the first voter could loose the secrecy of its vote. It seems that, at least,
the trusted authority should destroy the bootstrap votes once the checking has
been performed.

\section{Residual Problems}

As any paper trail or paper receipt protocol, the proposed protocol does not
prevent the machine from generating fraudulent votes that does not correspond
to any actual voter or bootstrap vote. Fraudulent receipts could be generated
and given to someone who could go in front of a court with them. However, in
that case, the number of votes would be greater than the number of signings.
One must then subtract the surplus from the result of the winner.

Concerning the secrecy of votes, one can imagine that somebody
forces many voters to show her their receipts. If she could collect many
receipts and she knew the order in which the votes were cast, she would still be
incapable of having the proof that a citizen voted in a certain way, but she
could possibly know that someone did not vote for a candidate, because
the ID associated with the vote would have already appeared in a previous
receipt. She could then retaliate. To avoid this
problem, it seems necessary to generate as many bootstrap votes as there are
registered voters, so that all the pairings between an ID and a choice
be unique. However, this solution could make it possible for the trusted
authority to break the secrecy of votes since it would be able to recognize
the actual votes among the bootstrap votes. Thus, this solution seems to bring
more risks than benefits.

At last, it should be noted that the proposed protocol is no solution
against some attacks targeting the secrecy of votes. For instance, it does not
prevent the analysis of electromagnetic radiations, the
introduction of a small video camera in the polling booth, such as those used
in cell phones, or the reconstruction of the order in which the voters cast
their ballot.

\section{Conclusion}

In some countries such as France, the voters enter the polling booth with as
many ballots as there are candidates, each ballot being marked for one
candidate. This protects the secrecy of votes. Here, we propose to turn this
procedure around by somehow giving to each voter as many
ballots as there are candidates, not before \textit{entering} the polling
booth, but before \textit{exiting} it.

The proposed protocol does not trust the software, as
any problem can be detected and corrected by the voters. Thus, it seems to be
``software independent''.~\cite{Riv0607} It is
moreover remarkable that the machines are essential to this protocol,
which cannot be manually implemented. However, some problems related to
the secrecy of votes remain and seem difficult to overcome completely. The
question then is
whether one is ready to trade a minimal risk for the secrecy against the
possibility to verify that one's vote has actually been taken into account.

\bigskip
\bigskip
\bigskip

\bibliographystyle{plain}

\end{document}